# Finite Element Model Updating Using Fish School Search Optimization Method

Boulkabeit I [*], Mthembu L [*] Marwala T [*] and De Lima Neto, F [†]

[*]The Centre For Intelligent System Modelling (CISM), Electrical and Electronics Engineering Department, University of Johannesburg, PO Box 524, Auckland Park 2006, South Africa, e-mail: iboulkaibet@student.uj.ac.za

[†] Polytechnic School of Pernambuco, University of Pernambuco, Recife, Brazil, e-mail: fbln@ecomp.poli.br

*Abstract*—**A recent nature inspired optimization algorithm, Fish School Search (FSS) is applied to the finite element model (FEM) updating problem. This method is tested on a GARTEUR SM-AG19 aeroplane structure. The results of this algorithm are compared with two other metaheuristic algorithms; Genetic Algorithm (GA) and Particle Swarm Optimization (PSO). It is observed that on average, the FSS and PSO algorithms give more accurate results than the GA. A minor modification to the FSS is proposed. This modification improves the performance of FSS on the FEM updating problem which has a constrained search space.**

## 1. INTRODUCTION

A finite element model (FEM) is a numerical method used to provide approximate solutions to the analysis of complex engineering systems [1, 2, 20]. In mechanical engineering, the FEM method is often used for computing displacements, stresses and strains in structures under different input conditions. In electrical engineering it might be used to model systems that experience electromagnetic fields. However, FEMs only produce accurate solutions to simple engineering problems. In complex systems, FEM results are different from those obtained from physical experiments [3, 4, 20]. These differences can be the result of; the approximations made in the geometric construction the system model and/or the description of some system parameters e.g. joints. Therefore the initial FE model needs to be automatically and intelligently updated so as to match the measured experimental data.

There are currently two main approaches to FEM updating; direct and indirect methods [1, 6]. In the direct method the measured data are directly equated to the FEM output. This naturally constrains the updating to FE system matrices (mass, stiffness etc) only. In the indirect (or iterative) approach the FEM outputs are not constrained to equal the measured data thus allowing both the system matrices and the model output to be variable. The updating problem is then how to minimise the difference between the measured data and the FEM output, given these "free" variables i.e. this is a classic optimization problem.

Another way of looking at this model updating problem is that we know what the model results should be, the problem is to identify the system that generate the results. So FEM updating can also be seen as a system identification problem which is made more difficult by the often large number of uncertain parameters.

Furthermore in FEM unlike in artificial models, the problem under consideration is a real physical system. Thus the updated parameters have physical meaning. This places significant limits on possible uncertain-parameter update values. This is one of the main criticisms of direct updating methods discussed above [1].

In the recent years, nature inspired optimization techniques, commonly referred to as metaheuristic, have shown promising results in systems identification research [5]. Many optimization methods such as Simulated annealing (SA), Genetic algorithm (GA) and Particle Swarm optimization (PSO) have been used for FEM updating [6, 7, 8]. In this paper we compare the performance of the recently introduced Fish School Search (FSS) algorithm [10] to the performance of the former algorithms on the FEM updating problem.

## 2. EVOLUTION SEARCH

### 2.1 Introduction

All nature-inspired search algorithms are population based and have the same structure [9]. They all have a population of individuals which move around the problem space searching for the solution to the problem. The dimension of the search space is normally defined by the number of uncertain function/problem parameters. All the individuals are potential solutions to the problem at hand. The following generally occurs for the individuals during the search process in all the above algorithms:

- Individuals are randomly initiated, in the beginning there is no bias to any particular point in the search space;
- Individuals interact to varying degrees at a local (exploitation) and global (exploration) level;
- Individuals are all evaluated during the search process;

- There is a level of randomness at all stages of the search process.

The novelty between algorithms is mainly how the potential solutions search/move (or update) in the problem space and how they interact (and or communicate) to each other.

The next section briefly describes the different algorithms compared in this paper. In section three the modeled structure is described and in section four the simulation results are presented. Section five concludes the paper.

### 2.2 Particle Swarm Optimization (PSO)

PSO was introduced by Kennedy and Eberhart [7]. It is based on the foraging behavior of a flock of birds (particles). As a standard in all PSO algorithms [6, 7, 8] each individual (*i*) is composed of three vectors: its position in the search space (size *d*) is given by $x_i = (x_i^1, x_i^2, x_i^3, ......x_i^d)$, the best position it has found $p_i = (p_i^1, p_i^2, p_i^3, ......p_i^d)$ and its velocity $v_i = (v_i^1, v_i^2, v_i^3, ...... v_i^d)$. The position and the velocity are initialized randomly in the search space.

The particles move around the search space by updating their velocity and position vectors. For the current context each position on the search space is a candidate solution to the FEM updating problem. The velocity can be updated using various equations; in this paper we use the PSO version with the inertia weight not the constriction factor [7].

The velocity update equation for particle *i* is given by:

$$v_i^d = wv_i^d + c_1 r_1 (p_i^d - x_i^d) + c_2 r_2 (p_g^d - x_i^d) \quad (1)$$

where *w* is the inertia weight introduced to adjust the influence of the previous velocities on the optimization process. A large inertia weight value helps in global exploration, while a small one helps in searching within nearby areas. The inertia weight used in this paper is given by:

$$w = \frac{\max ter - iter}{\max iter} \quad (2)$$

The position vector is then updated using

$$x_i^d = x_i^d + v_i^d \quad (3)$$

The $r_1$ and $r_2$ are independent random numbers and $c_1$ and $c_2$ are the local and global influence factors respectively.

### 2.3 Genetic Algorithm

Perhaps, the Genetic Algorithm (GA) is one of the first metaheuristic (of the nature inspired algorithms) used for optimization [6, 9, 22]. The idea stems from the way biological genes (more specifically chromosomes) of parents are combined to produce better and better off-springs over generations.

In the current context a potential solution to the FEM updating problem is modeled as a vector of chromosomes. Each chromosome is an uncertain parameter of the function to be optimized. Again for the current FEM updating structure (detailed in §3) the problem space *d =8*. In GAs an individual's make-up ($x_i$) is described by the chromosome vector itself. Unlike in PSO and FSS where there is a defining equation for how an individual moves through the search space, in GAs that notion is achieved by selection, reproduction and mutation between and of individuals. This allows the algorithm to gradually converge to better solutions over generations.

### 2.4 Fish School Search (FSS)

This method was inspired from the behavior of a school of fish and was first thought by Bastos Filho and Lima-Neto in 2007, subsequently proposed in 2008 [10].

Similar to PSO and GA, the search process in FSS is carried out by a population of individuals, i.e. fish, which have a certain level of local and global behavior. The problem space is said to have food (possible solutions) scattered in different concentrations of the aquarium (i.e. search space) and the fish are considered to 'feed' at these different positions as they search.

In FSS, searching is referred to as swimming. This action is achieved via three effectors; individual movement, collective-influence movement and school performance (called collective-volitive) movement. The initial individual position is random. Thereafter the movement is given by:

$$x_i(t) = x_i(t-1) + rand(-1,1) step_{ind}(t-1) \quad (4)$$

where *rand* is a random number uniformly generated in the interval [-1,1] and $step_{ind}$ is an individual step size. The step size decreases linearly during the search process so as to exploit later positions in the search process. Before movement to any particular point is taken the fish evaluates whether the food in that direction is better than at its current position. The fish then feeds. Individuals in FSS 'store' their previous performance information via their weight $w_i(t)$.

The larger this value is the more successful the fish has been in searching for food. The weight value is constrained to vary between 1 and $w_{scale}$. All fish are initialized with a weight of half $w_{scale}$. The weight of each individual in the population is updated using Equation 5:

$$w_i(t) = w_i(t-1) + \frac{f[x_i(t)] - f[x_i(t-1)]}{\max\{| f[x_i(t)] - f[x_i(t-1)] |\}} \quad (5)$$

where *f* is the finite element model and $f(x_i)$ computes the fitness of each fish in the school. After the individual movements a weighted average of all the fish is calculated. This biases the future movement of fish towards those fish that had a successful previous movement. The resultant fish movement is calculated using:

$$x_i(t) = x_i(t-1) + \frac{\sum_{i=1}^{N} \Delta x_{ind} \{f[x_i(t)] - f[x_i(t-1)]\}}{\sum_{i=1}^{N} \{f[x_i(t)] - f[x_i(t-1)]\}} \quad (6)$$

where $\Delta x_{ind}$ is the individual displacement of the fish and $N$ is the number of fish in the school. This is the collective – influence effector. The fish complete the swim movement with the collective - volitive adjustment step. This step takes into account how the whole school is performing so far. This step requires the calculation of the average weighted school position called the barycenter at time *t*:

$$B(t) = \frac{\sum_{i=1}^{N} x_i(t) w_i(t)}{\sum_{i=1}^{N} x_i(t)} \quad (7)$$

The final fish position will depend on whether the weight of the school has increased or decreased, this is given by Equations 8 and 9 respectively. The Equation 8 will be evaluated in the case that the weight has been increased. However, the Equation 9 will be used when the weight decreases:

$$x_i(t) = x_i(t-1) - step_{vol} \, rand \, \frac{\{x_i(t-1) - B(t-1)\}}{dist(\{x_i(t-1), B(t-1)\})} \quad (8)$$

$$x_i(t) = x_i(t-1) + step_{vol} \, rand \, \frac{\{x_i(t-1) - B(t-1)\}}{dist(\{x_i(t-1), B(t-1)\})} \quad (9)$$

The parameter $rand$ is a random number uniformly generated in the interval [0,1]. The $step_{vol}$ is used to control the displacement from or to the barycenter and it is decreased linearly as the search proceeds. The $dist(x, y)$ is the Euclidian distance between *x* and *y*.

This final positioning based on the average school weight increasing or decreasing effectively implements the space exploitation or exploration concept respectively. The fish exploit a particular area of space if the school collectively gains weight (i.e. school radius reduce) or otherwise explores other areas by expanding away from the barycenter if the school looses weight.

*2.5 Fish School Search biased (FSSb)*

The original FSS algorithm is slightly modified by giving more influence to results that seems to be more promising. This was done by simply altering the feeding of fish. A variable $\beta$, which allows for selective allocation of more weight to better search results is introduced to Equation 5. Obviously, this parameter is problem specific. The new equation is then:

$$w_i(t) = w_i(t-1) + \beta \frac{f[x_i(t)] - f[x_i(t-1)]}{\max\{|f[x_i(t)] - f[x_i(t-1)]|\}} \quad (10)$$

where

$$\begin{cases} \beta = 1.5, \text{ if "i" is a local best position} \\ \beta = 2 \text{ if "i" is a global best position} \\ \beta = 1, \text{ otherwise} \end{cases} \quad (11)$$

## 3. MODELLED STRUCTURE AND FE MODEL

All the finite element modeling was simulated using version 6.2 of the Structural Dynamics Toolbox (SDT®) under MATLAB® environment. In this paper, a GARTEUR SM-AG19 aeroplane structure is used to investigate the optimization capability of the four algorithms. The GARTEUR SM-AG19 structure was used as a benchmark study by 12 members of the GARTEUR Structures and Materials Action Group 19 [13, 14, 19]. One of the aims of the study was to compare the different computational model updating procedures on a single common test structure [12, 13, 15, 16, 19, 18]. The benchmark study also allowed participants to test a single representative structure using their own test equipment. The experimental test data used in our analysis is data obtained by the University of Manchester (U MAN).

The above aeroplane has a length of 1.5 m and a width of 3m. The depth of the fuselage is 15cm with a thickness of 5cm. Figure 1 shows the FE model of the aeroplane. In our models all element materials are considered standard isotropic. The model elements are Euler–Bernoulli. The measured natural frequency (Hz) data is: 6.51, 16.37, 33.44, 33.97, 36.17, 49.41, 50.2, 55.61, 64.04, 69.39.

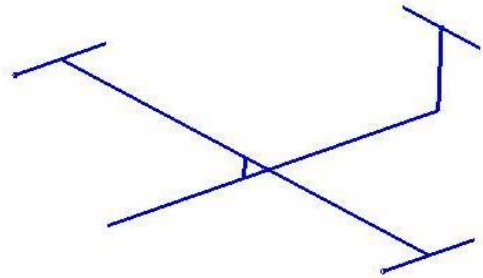

Figure 1 FEM Garteur Structure

The parameters to be updated are the right wing ($I_{min}, I_{max}, I_{tors}$), the left wing ($I_{min}, I_{max}, I_{tors}$), Vertical tail ($I_{min}$) and the overall structure's density ($\rho$). The search space is thus $x^d = 8$.

The initial position vector, $x_i(t)$, for all algorithms is given by E $=[\rho, VTP-I_{min}, L-I_{min}, L-I_{max}, L-I_{tors}, R-I_{min}, R-I_{max}, R-I_{tors}]$

Table 1 The initial vector position for all algorithms

| Parameter | $\rho$ $(kg/m^3)$ | $VTP-I_{min}$ $(10^{-9}m^4)$ | $L-I_{min}$ $(10^{-9}m^4)$ | $L-I_{max}$ $(10^{-7}m^4)$ |
|---|---|---|---|---|
| | 2700 | 8.3 | 8.3 | 8.3 |
| Parameter | $L-I_{tors}$ $(10^{-8}m^4)$ | $R-I_{min}$ $(10^{-9}m^4)$ | $R-I_{max}$ $(10^{-7}m^4)$ | $R-I_{tors}$ $(10^{-8}m^4)$ |
| | 4.0 | 8.3 | 8.3 | 4.0 |

## 4. SIMULATION RESULTS

In all simulations the number of iterations is set to 500.

### 4.1 PSO Settings

In the PSO algorithm, the population number is set to 20, and constants $c_1$ & $c_2$ are set to be 2. The particle position and velocity elements are bounded as shown in Table 2

Table 2 Particle position and velocity bounds

|  | $Max_{\_position}$ | $Min_{\_position}$ | $Max_{\_velocity}$ | $Min_{\_velocity}$ |
|---|---|---|---|---|
| $\rho$ | 3000 | 2000 | 10 | -10 |
| $VTP - I_{min}$ | $9.8 \times 10^{-9}$ | $7.3 \times 10^{-9}$ | $0.05 \times 10^{-9}$ | $-0.05 \times 10^{-9}$ |
| $L - I_{min}$ | $9.8 \times 10^{-9}$ | $7.3 \times 10^{-9}$ | $0.05 \times 10^{-9}$ | $-0.05 \times 10^{-9}$ |
| $L - I_{max}$ | $9.8 \times 10^{-7}$ | $7.3 \times 10^{-7}$ | $0.05 \times 10^{-7}$ | $-0.05 \times 10^{-7}$ |
| $R - I_{min}$ | $9.8 \times 10^{-9}$ | $7.3 \times 10^{-9}$ | $0.05 \times 10^{-9}$ | $-0.05 \times 10^{-9}$ |
| $L - I_{max}$ | $9.8 \times 10^{-7}$ | $7.3 \times 10^{-7}$ | $0.05 \times 10^{-7}$ | $-0.05 \times 10^{-7}$ |
| $L - I_{tors}$ | $5.5 \times 10^{-8}$ | $3 \times 10^{-8}$ | $0.05 \times 10^{-8}$ | $-0.05 \times 10^{-8}$ |
| $R - I_{tors}$ | $5.5 \times 10^{-8}$ | $3 \times 10^{-8}$ | $0.05 \times 10^{-8}$ | $-0.05 \times 10^{-8}$ |

### 4.2 GA settings

Table 3 shows the parameter setting used in the GA algorithm.

Table 3 GA parameter settings

| Population (*N*) | 20 |
|---|---|
| Mutation rate (*mutr*) | 0.2 |
| Selection Rate (*selcr*) | 0.5 |

As in the PSO algorithm the chromosomes in the genetic algorithm are bounded by the values shown in Table 2.

### 4.3 FSS settings

In FSS algorithm the fish population is 20. The population positions are bounded according to table 2. The $step_{ind}$ and $step_{vol}$ are a percentage of the search space amplitude and are bounded by two vector parameters according to Table 4:

Table 4 FSS parameter settings

| $step_{max}$ | $step_{min}$ | $stepvol_{max}$ | $stepvol_{min}$ |
|---|---|---|---|
| 30 | -30 | 0.08 | 0.06 |
| $0.08 \times 10^{-9}$ | $-0.08 \times 10^{-9}$ | 0.08 | 0.06 |
| $0.08 \times 10^{-7}$ | $-0.08 \times 10^{-7}$ | 0.08 | 0.06 |
| $0.08 \times 10^{-9}$ | $-0.08 \times 10^{-9}$ | 0.08 | 0.06 |
| $0.08 \times 10^{-7}$ | $-0.08 \times 10^{-7}$ | 0.08 | 0.06 |
| $0.08 \times 10^{-8}$ | $-0.08 \times 10^{-8}$ | 0.08 | 0.06 |
| $0.08 \times 10^{-8}$ | $-0.08 \times 10^{-8}$ | 0.08 | 0.06 |

The weight is bounded by: $w_{min} = 1$ and $w_{scale} = 250$.

Each algorithm is executed 30 times where the average solutions of these runs are presented in Table 5 and 6. Table 5 presents the initial value (the mean material or geometric values) of the update vector **E**, as well as the updated values obtained by each method. The updated parameters obtained by GA, PSO, FSS and FSSb techniques are close to the mean values i.e. they are physically realistic for the given structure.

In general, the updated vector obtained from GA, PSO and FSS algorithms are different from each other and this because of the way that each algorithm functions to generate the solution. The FSS and FSSb have very close update vector **E** where involving the global and the local best position slightly affected the updated vector **E**. PSO and FSS techniques give updated vectors that are more similar to each to those obtained by the GA algorithm.

Table 5 Initial and updated parameter values

|  | Initial **E** | **E** vector, GA Method | **E** vector, PSO Method | **E** vector, FSS Method | **E** vector, FSSb Method |
|---|---|---|---|---|---|
| $\rho$ | 2700 | 2573.7 | 2682.7 | 2687.1 | 2695.2 |
| $I_{x2}$ | $8.3 \times 10^{-9}$ | $7.6335 \times 10^{-9}$ | $9.0264 \times 10^{-9}$ | $9.1 \times 10^{-9}$ | $9.1 \times 10^{-9}$ |
| $I_{x2}$ | $8.3 \times 10^{-9}$ | $9.5178 \times 10^{-9}$ | $9.833 \times 10^{-9}$ | $9.8 \times 10^{-9}$ | $9.8 \times 10^{-9}$ |
| $I_{x2}$ | $8.3 \times 10^{-7}$ | $7.4915 \times 10^{-7}$ | $7.333 \times 10^{-7}$ | $7.333 \times 10^{-7}$ | $7.333 \times 10^{-7}$ |
| $I_{x2}$ | $8.3 \times 10^{-9}$ | $9.4841 \times 10^{-9}$ | $9.833 \times 10^{-9}$ | $9.8 \times 10^{-9}$ | $9.8 \times 10^{-9}$ |
| $I_{x2}$ | $8.3 \times 10^{-7}$ | $7.4632 \times 10^{-7}$ | $8.2885 \times 10^{-7}$ | $8.329 \times 10^{-7}$ | $8.372 \times 10^{-7}$ |
| $I_{x2}$ | $4 \times 10^{-8}$ | $3.6519 \times 10^{-8}$ | $3.6708 \times 10^{-8}$ | $3.69 \times 10^{-8}$ | $3.369 \times 10^{-8}$ |
| $I_{x2}$ | $4 \times 10^{-8}$ | $3.918 \times 10^{-8}$ | $3.6815 \times 10^{-8}$ | $3.65 \times 10^{-8}$ | $3.972 \times 10^{-8}$ |

Table 6 shows the output errors of the different optimization algorithms. In general, the results show that the updated FEM natural frequencies are better than the initial FEM for all methods (see the total (T) error in Table 6). The error between the second measured natural frequency and that of the initial model was 6.303%. With the GA method this error was reduced to 0.915% and by implementing the PSO it was reduced to 0.763%, and when the FSS and FSSb are implemented, the error reduced to 0.727% and 0.651% respectively. The same comment can be made for the third, fourth, seventh, eighth, ninth and the tenth natural frequencies.

The PSO, FSS and FSSb give results that are better than the GA algorithm. On the other hand, GA converges faster to a local minimum, see figure 2. FSS and FSSb converge within the first 15 iterations. PSO, FSS and FSSb algorithms have almost the same asymptotic point (around 70 to 75 iterations). There is no significant difference between the total errors obtained for the PSO, FSS and FSSb [21].

Table 6 Natural Frequencies and Errors when GA, PSO, FSS and FSSb methods are used to update the FEM

| Mode | Measured Frequency (Hz) | Initial FEM Frequency (Hz) | Error (%) | Frequencies GA Method (Hz) | Error (%) | Frequencies PSO Method (Hz) | Error (%) | Frequencies FSS Method (Hz) | Error (%) | Frequencies FSSb Method (Hz) | Error (%) |
|---|---|---|---|---|---|---|---|---|---|---|---|
| 1 | 6.51 | 5.726 | 12.047 | 6.247 | 4.037 | 6.237 | 4.195 | 6.232 | 4.266 | 6.224 | 4.396 |
| 2 | 16.37 | 15.338 | 6.303 | 16.22 | 0.915 | 16.495 | 0.763 | 16.489 | 0.727 | 16.77 | 0.651 |
| 3 | 33.44 | 32.457 | 2.939 | 33.086 | 1.057 | 33.306 | 0.399 | 33.278 | 0.486 | 33.350 | 0.269 |
| 4 | 33.97 | 35.323 | 3.984 | 34.949 | 2.882 | 34.154 | 0.542 | 34.127 | 0.461 | 34.222 | 0.742 |
| 5 | 36.17 | 36.020 | 0.414 | 36.284 | 0.316 | 35.908 | 0.724 | 35.896 | 0.759 | 35.925 | 0.678 |
| 6 | 49.41 | 44.992 | 8.941 | 49.05 | 0.728 | 48.839 | 1.156 | 48.799 | 1.237 | 48.730 | 1.377 |
| 7 | 50.20 | 54.685 | 8.934 | 53.964 | 7.499 | 52.938 | 5.455 | 52.903 | 5.384 | 52.834 | 5.248 |
| 8 | 55.61 | 55.753 | 0.257 | 54.603 | 1.811 | 55.512 | 0.176 | 55.574 | 0.065 | 55.603 | 0.012 |
| 9 | 64.04 | 60.021 | 6.276 | 63.695 | 0.538 | 64.16 | 0.187 | 64.130 | 0.140 | 64.068 | 0.043 |
| 10 | 69.39 | 68.745 | 0.929 | 70.326 | 1.349 | 68.922 | 0.675 | 68.868 | 0.752 | 68.768 | 0.859 |
| T error % | ______ | ______ | 51.024 | ______ | 21.132 | ______ | 14.272 | ______ | 14.277 | ______ | 14.275 |

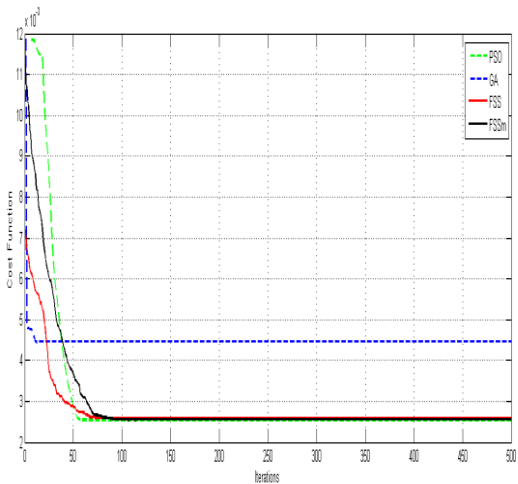

Figure 2 FEM cost function vs. iterations

It is difficult to make a decisive conclusion from Table 6 and Figure 2 when comparing the FSS, FSSb and PSO algorithms. The difference between the algorithm results is on the different natural modes. The FSS algorithm has smaller errors than PSO in five of the modes and PSO has smaller errors than FSS for a different set of five modes. On average the total error between these algorithms is similar as shown in figure 2.

The modified FSS algorithm, FSSb, outrank PSO on six out of the 10 modes. For example the error of the second mode was reduced to 0.763 % by the PSO algorithm while the FSSb method reduced it to 0.651 %. The same comment can be made for the third, fifth, seventh, eighth and ninth natural frequencies. Figure 3 shows a magnified plot of the cost function at higher iteration counts.

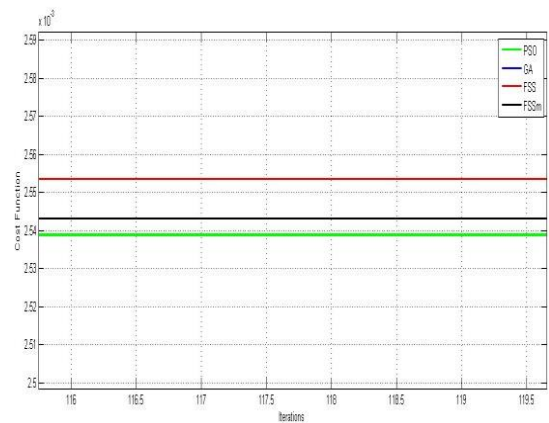

Figure 3 Magnified FEM cost function vs. iterations

Figure 3 also shows that the total error of the PSO algorithm is slightly better than both FSS and FSSb algorithms, even though as commented it does not converge as fast. The monification proposed in FSSb seemed to slightly improve results over

"vanilla" FSS. The algorithm errors do not improve after the magnified region.

*5. CONCLUSION*

In this paper the finite element model updating problem is addressed using a recently proposed nature inspired optimization algorithm. Three of these metaheuristic methods; the genetic algorithm (GA), particle swarm optimization (PSO) and fish school search (FSS) algorithms are tested on the updating of the GARTEUR SM-AG19 FEM structure. A slight modification to the recent FSS algorithm is proposed and tested; we called it FSSb ('b' for bias).

The simulation results show that the FSS, FSSb and the PSO algorithms give more accurate results than those obtained by the GA algorithm. On the other hand, the GA algorithm has a faster convergence rate than the former although it prematurely converges. The modification on FSS algorithm improves the results of the FEM over the PSO algorithm. However, the PSO algorithm gives slightly better total error than those obtained by both FSS and FSSb algorithms.

Further work will consider the differences between the above methods and the density based FSS algorithm (dFSS) [21]. Our preliminary analysis reveals that the FSSb algorithm may be more suited for applications that have a discretized search space, such as in FEM. But this has to be further verified in other domains.